\documentclass[runningheads,10pt]{llncs}
\usepackage[T1]{fontenc}
\usepackage{graphicx}
\usepackage{caption}
\usepackage{mathtools}
\usepackage{amssymb}
\usepackage{enumitem}
\usepackage{multicol}
\usepackage{appendix}
\usepackage{subcaption}
\usepackage[dvipsnames]{xcolor}
\usepackage{tikz}
\usepackage{pgfplots}
\usetikzlibrary{patterns}
\pgfplotsset{compat=1.18}

\providecolor{bluetwo}{RGB}{65,105,225}
\providecolor{bluethree}{RGB}{70,130,180}

\usepackage{tcolorbox}
\usepackage{booktabs}
\usepackage{hyperref}
\usepackage{cleveref}
\usepackage{dsfont}
\usepackage{xurl}
\usepackage{xcolor}
\usepackage{algorithm}
\usepackage{algorithmic}

\DeclareMathOperator*{\argmin}{arg\,min}

\begin{document}
\title{Agentic Incident Response through Digital Twin-Enhanced Multiscale Planning}
\titlerunning{Agentic Incident Response Multiscale Planning}

\author{Yiran Gao\inst{1} \and Tao Li\inst{1} \and Kim Hammar\inst{2}}

\institute{City University of Hong Kong, Hong Kong SAR\\ \email{gaoyiran525@gmail.com, li.tao@cityu.edu.hk} \and Imperial College London, United Kingdom\\ \email{k.hammar@imperial.ac.uk}}

\authorrunning{Gao, Y., Li, T., and Hammar, K.}
\maketitle  

\begin{abstract}
Incident response is currently managed by security operators using predefined playbooks, resulting in slow, labor-intensive security decision-making processes. Consequently, there is a growing need for automated incident response planning.  Decision-theoretic approaches based on control, optimization, and reinforcement learning have been proposed to automate such planning tasks with well-grounded approaches, yet most of which, while guaranteeing strong performance, are limited to abstract models and cannot be directly applied to operational systems. A promising approach to mitigate this limitation is to use the security knowledge embedded in large language models (LLMs) to develop agentic response systems. However, current agentic approaches rely on repeated invocations of the LLM to generate a response plan, which is unreliable and limits the planning horizon due to hallucination. In this paper, we develop a principled LLM-based planning method by combining decision-theoretic planning with LLM-generated response commands. The proposed agentic incident response approach uses a rollout planner to compute a high-level response strategy that allocates security resources (the tactical scale), which is then translated into executable commands by a lightweight LLM agent (the operational scale). Within this architecture, we use a digital twin that supports tactical planning through simulation and operational execution through emulation. Across three attack scenarios, our agentic approach reduces recovery execution time by 15.1\% on average and increases the recovery rate by 33.6\% over frontier LLM baselines.

\end{abstract}

\section{Introduction}
Incident response refers to the coordinated actions taken to contain, mitigate, and recover from cyberattacks. Today, incident response is largely a manual process carried out by security operators. Though this approach can be effective, it is often slow and requires specialized skills. For instance, a recent report by IBM indicates that 60\% of the surveyed organizations take more than 100 days to respond and recover from security incidents in networked systems \cite{ibm-report}.

To address the limitation of manual security operations, autonomous cyber defense (ACD) has emerged as a promising approach for developing agents that can respond to attacks without human intervention \cite{burnap25acd-survey,reddi23rl4sec,li2025agentic}. For incident response planning, existing ACD methods typically rely on abstract decision-making models and simulators; see, e.g., \cite{zhu13gamesec}. By converting the networked system into an abstract simulation (e.g., a Markov decision process \cite{erik18pomdp-defense,tao25deception,xinqi26deplay-rl}), these methods make the planning problem tractable and leverage control and optimization, game theory, and reinforcement learning (RL) to compute optimal plans. Yet, this abstraction also limits the practical applicability of the resulting response plans. In particular, response plans produced by current ACD methods operate at the \textit{tactical scale}: they prescribe high-level defensive actions (e.g., defend this host) without specifying the implementation at the \textit{operational scale}.

A promising approach to close this gap is to use a large language model (LLM) to automatically generate an operational response plan, i.e., one that includes executable commands for a networked system. Unlike abstract decision-theoretic agents, LLM agents can process large volumes of system logs and generate executable system commands \cite{llm4sec-review25}. However, current LLM-based approaches rely mostly on prompt engineering of general-purpose LLMs without principled planning algorithms, which is unreliable and prone to hallucinations \cite{hamoun25llm-acd,cardenas25llm4acd,guo25ircopilot}.
\begin{figure}
\vspace{-1em}
  \centering
\includegraphics[width=0.9\linewidth]{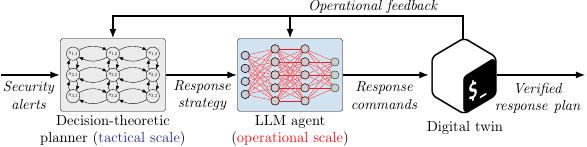}   
 \caption{Illustration of our multiscale approach to agentic incident response planning. Tactical scale: We employ a decision-theoretic planner to generate a high-level response strategy. Operational scale: We use a Large Language Model (LLM) agent to translate the high-level strategy into executable commands and verify them in a digital twin.}
\label{fig:multiscale}
\vspace{-1em}
\end{figure}
To address these limitations and bridge the gap between the tactical and operational scales, we propose combining decision-theoretic planning with LLM-based response generation; see Fig.~\ref{fig:multiscale}. On the tactical scale, we use the lookahead rollout method, a major RL approach \cite{tao23cola,bertsekas24rollout}, to compute a high-level response strategy based on an abstract model. This strategy is then fed to an LLM agent, which operates at the operational scale by grounding it in the system context and translating it into an executable response plan.  This separation allows the planner to provide tactical guidance, while the LLM handles the system-specific details required for execution.

Within this multiscale planning framework, we use a digital twin (i.e., a virtual replica of the system affected by the incident) that supports tactical planning through simulation and operational execution through emulation \cite{hammarcsle}. In particular, simulation enables computationally efficient planning at the tactical scale by abstracting system details and simulating attack progression across the network, while emulation enables operational-scale response action verification by allowing testing of the generated response plan before deployment. 

We implement our multiscale architecture on a testbed network comprising five servers (Fig~\ref{fig:dt}) and use it to recover the networked system under three attack scenarios that exploit diverse vulnerabilities across servers. In contrast to LLM agents proposed in prior work \cite{hamoun25llm-acd,cardenas25llm4acd,guo25ircopilot}, our implementation can be deployed locally and does not rely on an external LLM provider. In particular, we fine-tune the Deepseek-R1-14B LLM \cite{deepseek-r1} on an open-source dataset of $68,000$ incidents and their corresponding responses. Despite being so lightweight ($14$B parameters compared to $\approx 800$B), we show that our system outperforms frontier LLMs by 15\% on average across the attack scenarios we evaluated, in terms of recovery time. Moreover, it outperforms frontier LLM baselines by {15.1\%} on average in recovery execution time, while maintaining a success rate over 90\% across distinct scenarios. Our contributions are as follows. 
\begin{itemize}
    \item We formulate the incident response as a factorized Markov decision process that separates the network-level, tactical-scale planning from the server-specific, operational-scale planning.
    \item We develop an agentic incident response framework that integrates decision-theoretic rollout planning at the tactical scale with LLM-based generation verified by a digital twin at the operational scale.
    \item We implement our architecture using a lightweight LLM and demonstrate that it outperforms the state of the art in recovery-action execution time and success rate for three multi-stage attacks.
\end{itemize}

\section{Related Works}
\noindent{\it\bfseries Decision-theoretic incident response.}
Since incident response can be viewed as a sequential decision-making process, there have been considerable and ongoing efforts to apply control, optimization, and, most recently, reinforcement learning (RL) methods to incident response. The general recipe, which can be traced back to \cite{kreidl04feedback-defense}, is to first model the response process as a discrete-time control system, where the system state encapsulates the network security posture and control actions are security responses; examples include the Markov decision process (MDP) \cite{burnap25acd-survey} and the Markov game \cite{alpcan10book,tao24ddztd}, among others. Then, the optimal response plan corresponds to the optimal policy of the sequential decision-making models. 

Recent developments \cite{iannucci18mdp-cyber,miehling22control-cyber} have actively engaged with robust control and optimal control to achieve optimal response planning in the offline design, while online learning \cite{kim-tao25col,kim-tao25quantization}, meta learning \cite{tao23ztd} and multi-agent learning \cite{tao24col,tao22info} have also been explored recently for online adaptive planning. Related to control-theoretic response planning, RL has emerged as a promising paradigm for data-driven optimal control when closed-form modeling of the network system is unavailable \cite{tao22confluence}. However, RL still relies on an accurate simulator of the target system, which is rarely available in practice.

\noindent{\it \bfseries LLM-based approaches for incident response.} 
A promising approach to address the drawbacks of decision-theoretic approaches is to use large language models (LLMs) to automatically generate effective response actions from system logs. This approach is not limited to a predefined set of actions and eliminates the need for a simulator. Early studies in this direction include \cite{hamoun25llm-acd,guo25ircopilot,tao25deception,wang2026cybergym}. These approaches can be separated into two categories: prompt-based LLM orchestration and LLM-RL hybrid approaches. 

The first category decomposes incident response into several subtasks and develops tailored prompts for LLMs for tackling each task \cite{hamoun25llm-acd,guo25ircopilot}. While these works report encouraging results, they have three key limitations: they do not leverage principled planning techniques, they rely on extensive prompt tuning, and most require uploading incident data to external LLM providers.

The second category addresses some of these limitations by combining RL and LLM agents, where RL agents supervise the LLM generation \cite{keman24rl-mentor}, LLM agents augment RL agents through knowledge sharing \cite{lopes24hybrid-ai}, and two agents communicate with each other \cite{cardenas25llm4acd}. Despite the different nature of agentic interactions, these works require additional RL training and LLM-RL joint operations in a simulated environment, leaving a gap between simulation and practical implementation. Most relevant to our work is a line of recent work on integrating decision-theoretic planning with LLM generation \cite{kim26ndss,yiran26e2e-ir}. However, these approaches use planning methods, e.g., rollout and Monte Carlo tree search, to guide LLM-generated operational response actions.  

\noindent{\it \bfseries Digital twins in cybersecurity.}
The concept of digital twins (DTs) originated in the manufacturing, aviation, and physical AI sectors \cite{grieves2016digital,tao25dt-pirl}, where a digital replica runs in parallel with the physical process, offering real-time situational awareness \cite{grieves2016digital,tao25dima}. In the context of cybersecurity, DTs provide two main kinds of functionalities: emulation and simulation; see e.g., \cite{dietz21dt-forensics,weippl19dt-aware,maclaughlin22dt-ml,pernul20dt,colomo20dt-survery,repetto26cdt,pernul20dt-soc}. Emulation aims to reproduce the target system’s functions and timing behavior in a virtual replica. Such a virtual replica can be used for digital forensics before real-world forensics takes place \cite{dietz21dt-forensics} and for visualizing vulnerabilities \cite{weippl19dt-aware}. More importantly, it provides a controlled environment for virtual operations, with outcomes that can be used to optimize operations in the target system \cite{repetto26cdt}, such as event management \cite{pernul20dt} and incident prediction \cite{colomo20dt-survery}. In contrast, simulation is a lightweight abstraction that models selected system aspects to explore security scenarios without reproducing the behavior of the full target system \cite{hammarcsle}.

\noindent{\it \bfseries Novelty of our approach.} To our knowledge, we are the first to combine tactical planning with operational response generation, whereas prior work focuses on either tactical planning or operational response generation in isolation. Compared with the most relevant works \cite{kim26ndss,yiran26e2e-ir}, our work features integrated tactical-operational planning that goes beyond LLM-based planning and is augmented with digital-twin verification. Moreover, unlike prior work that typically uses digital twins for either simulation or emulation, our method uses both: simulation for tactical-scale planning and emulation for operational-scale verification. Finally, by fine-tuning a local LLM, as in our previous work \cite{yiran26e2e-ir}, our approach is more lightweight and reduces dependence on external LLM providers.

\section{Preliminaries}
\label{sec:pre}
This section presents basic definitions that will be used in the subsequent section covering our methodology. We first describe the main stages of incident response and then review partially observable Markov decision processes (POMDPs) as a formal model for planning under partial observability.

\noindent\textit{\textbf{Incident Response Planning.}}
Incident response involves selecting a sequence of actions to restore a networked system to a secure, operational state after a cyberattack. These actions should analyze the scope of the attack, secure forensic evidence, contain and evict the attacker, harden the system to prevent recurrence, and restore critical services. Examples of response actions include redirecting network flows, updating access control policies, patching vulnerabilities, shutting down compromised systems, and restarting operational services.

We model the system affected by the incident using a graph $\mathcal{G}=\langle \mathcal{V}, \mathcal{E}\rangle$. The node set $\mathcal{V}\triangleq\{1, 2, \ldots, N\}$ includes $N$ components, and their interconnections are denoted by the edge set $\mathcal{E}$. The system operator, which we refer to as the defender, monitors the system using infrastructure statistics from an intrusion detection system (IDS). Once an attack has been detected, the defender's task is to plan a sequence of actions to restore the system to a secure, operational state as quickly as possible.

Following the MITRE D3FEND taxonomy \cite{d3fend}, we divide the incident response process into the following six stages.
\begin{enumerate}[nosep]
    \item \textit{\textbf{Containment}}: isolating the attack and preventing it from spreading to other connected components.
    \item \textit{\textbf{Assessment}}: identifying the scope and severity of the attack.
    \item \textit{\textbf{Preservation}}: preserving forensic evidence for analysis.
    \item \textit{\textbf{Eviction}}: revoking the attacker's access to the system.
    \item \textit{\textbf{Hardening}}: patching vulnerabilities and hardening the system to prevent recurrence of the attack.
    \item \textit{\textbf{Restoration}}: restoring services and user access.
\end{enumerate}
The goal of the defender is to find a sequence of actions that drives the affected system through these response stages as quickly as possible while minimizing operational costs. A key challenge in selecting such actions is that the available information about the attack is often limited to partial indicators of compromise, e.g., IDS alerts. Moreover, the attacker's tactics are generally unknown.

\noindent\textit{\textbf{Partially Observable Markov Decision Processes.}}
Given the partial observability of the system's security state, we formulate incident response planning as a partially observable Markov decision process (POMDP). We briefly review the formalism of POMDPs for a general use case, deferring the details of our multiscale planning system model to the next section.

A POMDP evolves over time steps $ t=0,1,2, \ldots$. At each time step $t$, the system's security status is represented by the unobservable state variable $s_t$. The defender has access to an observation $o_t$ that represents system metrics correlated with the recovery state (e.g., log files and IDS alerts). Specifically, the correlation between states and observations is modeled by an observation kernel $O(o_t \mid s_t)$, which defines the probability of observing $o_t$ in state $s_t$. At each time step $t$, a response strategy $\pi$ prescribes an action $a_t=\pi(o_{0:t})$ that influences the system's state evolution according to the Markov transition kernel $P_\theta$, where $\theta$ denotes the attacker's tactics and $P_\theta(s_{t+1}\mid s_t, a_t)$ specifies the probability of transitioning to state $s_{t+1}$ when executing response action $a_t$ in state $s_t$. We assume the existence of an absorbing terminal state $s_T$ such that $P_\theta(s_T\mid s_T, a)=1$ for all actions. This state models the operating conditions when the system has fully recovered and remains operational thereafter.

The time required to implement each response action may vary. For example, isolating a compromised host may take a few seconds, while performing forensic analysis of affected systems may last several hours. We model this time through a cost function $c$, where $c(s_t, a_t)$ represents the time to execute response action $a_t$ in state $s_t$. Naturally, $c(s_T, a)=0$ for all response actions $a$, since the terminal state does not require any response. Given this cost function, the problem of minimizing the recovery time can be expressed as $\min_{\pi}\mathbb{E}_{P_\theta,\pi}\left[ \sum_{t=0}^H c(s_t,a_t)\right]$, where $H$ is a finite time horizon. 

\section{Formalizing the Incident Response Use Case}\label{sec:system_model}

We formulate incident response as a planning process at two levels: (\textit{i}) tactical planning for allocating security resources to system components; and (\textit{ii}) operational planning of recovery actions for those components. To capture this two-level planning mathematically, we model incident response as a factorized POMDP where the state $s_t$ is decomposed into a security posture (tactical level) and a local recovery status (operational level), as detailed below.

\noindent\textit{\textbf{System states.}}
Consider an IT infrastructure with $N$ components. Each component can be in two global states: safe ($0$) or compromised ($1$). Consequently, the \textit{global} security state is the Boolean vector: $g_t=(g^1_t, \ldots, g_t^N),  g^k_t\in \{0, 1\}$. Moreover, each component $k$ is associated with a \textit{local} recovery state, which is defined as the 6-dimensional Boolean vector
\begin{align*}
\ell^k_t=(\ell^{k(c)}_t, \ell^{k(a)}_t, \ell^{k(p)}_t, \ell^{k(e)}_t, \ell^{k(h)}_t, \ell^{k(r)}_t), && \ell^k_t\in \{0, 1\}^6,
\end{align*}
where the $l$th entry indicates whether the $l$th response stage in \Cref{sec:pre} has been completed. For example, $\ell^{k(c)}_t=1$ if the attack has been contained; $\ell^{k(c)}_t=0$ otherwise. Hence, the system state $s_t$ is obtained by concatenating the global and local states, i.e., $s_t=((g_t^k,\ell_t^k)_{k\in [N]})$, where $[N] \triangleq \{1,2,\ldots, N\}$.

\noindent\textit{\textbf{Beliefs.}}
As discussed in \Cref{sec:pre}, the state is unobservable to the defender who needs to form a belief of the state using partial observations $o_{0:t}$. Mathematically, a belief is a probability distribution over possible states. Due to the binary structure of $s_t$, the probabilistic belief admits the representation $b_t=(b_t^g, (b_t^k)_{k\in [N]})$, where $b_t^g\in [0,1]^N$ and $b_t^g(k)$ indicates the probability of the $k$-th component being safe. Similarly, $b_t^{k(i)}$ indicates the probability that the $i$-th response stage has been completed for component $k$. One important assumption we impose is that each server's state is independent of the rest, and hence, $b_t^g$ is a product of Bernoulli distributions. This assumption greatly simplifies the belief space. 

\noindent\textit{\textbf{Actions.}}
Since tactical and operational planning are distinct, we model a response action as a tuple $a_t=(a_t^g, a_t^{\ell})$, where $a_t^g$ is the action on the tactical level and $a_t^{\ell}$ is the action at the operational level. Specifically, the tactical action $a_t^g$ is a permutation of $\{1,2, \ldots, N\}$ indicating the response priority over the networked components, where $a_t^g(k)$ is the priority of system component $k$. Similarly, the operational action $a_t^{\ell}$ corresponds to the recovery action (e.g., a system command) applied to the prioritized system component.

\noindent\textit{\textbf{Costs.}}
Given these definitions of the security state $s_t$ and response action $a_t$, we define the cost function as $
c(s_t,a_t)=c^g(g_t, a_t^g)+c^\ell(\ell_t, a^\ell_t)$,
where $c^\ell$ models the time required to execute the selected recovery action on the chosen component and $c^g$ captures the cost of delaying recovery for the remaining components. This delay matters because unattended compromised components may vary in criticality and importance. To model this importance, we define $r(g_t)\in \mathbb{R}^N$ to be a vector of importance weights, where $r_k(g_t)$ is the importance of component $k$ in global state $g_t$ and $\sum_{k\in[N]} r_k(g_t)=1$. Given these importance weights, we define $c^g(g_t,a_t^g) = \lambda \left(\sum_{j\neq k} r_j(g_t)a_t^g(j)\right) c^\ell(\ell_t,a_t^\ell)$,
where $k$ is the component selected for recovery. Thus, the global cost penalizes recovery orders that leave important or risky components waiting. The parameter $\lambda\in(0,1]$ controls the relative weight of the delay cost compared with the execution time cost.
We illustrate this cost definition through the following example.
\begin{figure}[!h]
  \centering
  \scalebox{1}{
\begin{tikzpicture}[
    server/.style={circle, draw, thick, minimum size=5mm, font=\scriptsize, inner sep=0pt},
    compromised/.style={server, fill=red!18},
    safe/.style={server, fill=green!12},
    recovered/.style={server, fill=gray!20, dashed},
    urgent/.style={server, fill=orange!35, very thick},
    edge/.style={->, semithick, >=stealth},
    inactive/.style={->, semithick, dashed, gray, >=stealth},
    labelbox/.style={align=center, font=\scriptsize},
    el/.style={font=\tiny, inner sep=1pt}
]
\node[compromised, minimum size=3mm] (L1) at (-1.55,1.5) {};
\node[anchor=west,font=\scriptsize,inner sep=2pt] at (L1.east) {Compromised};
\node[safe, minimum size=3mm] (L2) at (0.65,1.5) {};
\node[anchor=west,font=\scriptsize,inner sep=2pt] at (L2.east) {Safe};
\node[recovered, minimum size=3mm] (L3) at (1.95,1.5) {};
\node[anchor=west,font=\scriptsize,inner sep=2pt] at (L3.east) {Recovered};
\node[urgent, minimum size=3mm] (L4) at (3.8,1.5) {};
\node[anchor=west,font=\scriptsize,inner sep=2pt] at (L4.east) {Top priority};

\node[labelbox] at (0,0.8) {\textbf{Step $t$}: before\\recovering central server};
\node[urgent] (A1) at (0,0) {$v_1$};
\node[compromised] (A2) at (-0.85,-0.9) {$v_2$};
\node[compromised] (A3) at (0.85,-0.9) {$v_3$};
\node[safe] (A4) at (-1.35,-1.8) {$v_4$};
\node[safe] (A5) at (-0.48,-1.8) {$v_5$};
\node[safe] (A6) at (0.48,-1.8) {$v_6$};
\node[safe] (A7) at (1.35,-1.8) {$v_7$};
\draw[edge] (A1) -- node[el,left] {$0.8$} (A2);
\draw[edge] (A1) -- node[el,right] {$0.7$} (A3);
\draw[edge] (A2) -- node[el,left] {$0.4$} (A4);
\draw[edge] (A2) -- node[el,right] {$0.3$} (A5);
\draw[edge] (A3) -- node[el,left] {$0.7$} (A6);
\draw[edge] (A3) -- node[el,right] {$0.6$} (A7);
\node[labelbox] at (0,-2.45) {Priority order: $a_t^g(v_1){=}1,$\\$a_t^g(v_2){=}2,\ a_t^g(v_3){=}3$};

\begin{scope}[xshift=3.8cm]
\node[labelbox] at (0,0.8) {\textbf{Step $t{+}1$}: after\\recovering central server};
\node[recovered] (B1) at (0,0) {$v_1$};
\node[compromised] (B2) at (-0.85,-0.9) {$v_2$};
\node[urgent] (B3) at (0.85,-0.9) {$v_3$};
\node[safe] (B4) at (-1.35,-1.8) {$v_4$};
\node[safe] (B5) at (-0.48,-1.8) {$v_5$};
\node[safe] (B6) at (0.48,-1.8) {$v_6$};
\node[safe] (B7) at (1.35,-1.8) {$v_7$};
\draw[inactive] (B1) -- node[el,left] {$0.8$} (B2);
\draw[inactive] (B1) -- node[el,right] {$0.7$} (B3);
\draw[edge] (B2) -- node[el,left] {$0.4$} (B4);
\draw[edge] (B2) -- node[el,right] {$0.3$} (B5);
\draw[edge] (B3) -- node[el,left] {$0.7$} (B6);
\draw[edge] (B3) -- node[el,right] {$0.6$} (B7);
\node[labelbox] at (0,-2.45) {Updated priority: $a_{t+1}^g(v_3){=}1,$\\$a_{t+1}^g(v_2){=}2$};
\end{scope}
\end{tikzpicture}
}
\caption{Representation of the global state $g_t$ and the priority weights $r(g_t)$ as an attack graph; cf.~Example~\ref{exam:state-dependent-factor}. Nodes represent system components $i\in \{1,\hdots,N\}$, which are colored green if $g^i_t=0$ and red otherwise. Edges represent attack steps with the associated probabilities. The top priority node is indicated in orange. The left and right graphs represent the priorities at time $t$ and $t+1$, respectively.}
\label{fig:state-dependent-priority}
\end{figure}
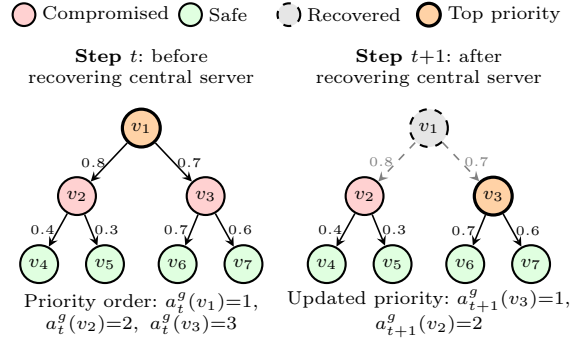

\begin{example}
\label{exam:state-dependent-factor}
The global state $g_t$ and the priority weights $r(g_t)$ can be illustrated using an attack graph, where nodes represent system components, directed edges represent attack steps, and edge weights represent probabilities of attack steps; see Fig.~\ref{fig:state-dependent-priority}. As shown in the left graph of Fig.~\ref{fig:state-dependent-priority}, system component $v_1$ receives the highest priority at step $t$ because it connects two compromised regions of the network. Recovering $v_1$ can therefore reduce the risk of further attack propagation. After component $v_1$ is recovered, the priorities among the remaining unrecovered components change, which results in a new attack graph with a different structure, as shown in the right graph of Fig.~\ref{fig:state-dependent-priority}.
\end{example}

\noindent\textit{\textbf{State dynamics.}} State transitions are coupled across the global and local scales. In particular, the local recovery action $a_t^{\ell}$ updates the local state  $\ell_t^k$ according to $a_t^g$, which determines whether the global state of that component (i.e., $g^k_t$) is updated or not. This coupling gives the factorized transition model $ P_\theta(s_{t+1}\mid s_t, a_t)=P^g_\theta(g_{t+1}\mid g_t, a^g_t)P_\theta^k(\ell^k_{t+1}\mid \ell^k_t, a^\ell_t)$, $a_t^g(k)=1$,
where $k$ denotes the component selected for recovery and $P_{\theta}^k$ models the progress of component $k$ through the response stages under action $a_t^{\ell}$. The global state transition follows from the outcome of the local action. Specifically, if component $k$ is compromised and its local state reaches the recovered state $\ell^k_T = (1,1,1,1,1,1)$, then its global state becomes safe ($g^k_{t+1}=0$). Otherwise, it remains compromised ($g^k_{t+1}=1$), as formally expressed below.
\begin{align*}
P_{\theta}^g(g^k_{t+1} = 0 \mid g^k_t = 1) = \mathds{1}\{\ell_{t+1}^{k} = \ell^k_T\},
P_{\theta}^g(g^k_{t+1} = 1 \mid g^k_t = 1) = \mathds{1}\{\ell_{t+1}^{k} \neq \ell^k_T\}.
\end{align*}
For all other components, the global state transition is governed by the attack graph, which models how the attack spreads from the compromised nodes. Specifically, let $g^{-k}_t$ denote the global state of all components except $k$, i.e., $g_{t+1}=(g^k_{t+1}, g^{-k}_{t+1})$. Then, $P_{\theta}^g(g_{t+1} \mid g_t, a^g_t) = P_{\theta}^{g}(g^k_{t+1} \mid g^k_t)P_{\theta}^{g}(g_{t+1}^{-k} \mid g^{-k}_t)$.

These definitions imply that the global and local transition kernels differ in both timescale and granularity. In particular, the global state $g^k_t$ changes only after component $k$ has completed several recovery steps and reached its terminal recovery state. Thus, global transitions describe network-level changes in compromise status, while local transitions capture the detailed recovery progress of an individual component, which motivates our multiscale planning approach.

\section{Agentic Multiscale Response Planning}
In this section, we present our method for incident response planning. It includes an offline stage and an online stage. In the offline stage, we fine-tune a lightweight LLM using incident response examples. In the online stage, we use the fine-tuned LLM and a digital twin to plan responses at two scales: tactical and operational. At the tactical scale, a decision-theoretic planner uses digital twin simulation to prioritize which component to recover next. At the operational scale, an LLM-based agent generates recovery actions for that component and verifies them through digital twin emulation. 

\subsection{Offline Fine-tuning of a Lightweight LLM}
We adapt the DeepSeek-R1-14B \cite{deepseek-r1} LLM to incident response by fine-tuning it on a labeled dataset of incident descriptions. We fine-tune three variants of the model, each corresponding to a different stage of the response workflow.

\noindent\textit{\textbf{Incident assessment fine-tuning.}}  We start by fine-tuning the LLM to infer likely attacker tactics and techniques from an incident description, such as system logs and security alerts. We conduct this fine-tuning by training the LLM on a labeled dataset of incident examples, denoted by $\mathcal{D}_{\text{incident}}=\{(\mathbf{x}^i, \mathbf{y}^i)\}_{i=1}^K$. Each input $\mathbf{x}^i$ contains an incident description (e.g., security alerts), while the target output $\mathbf{y}^i$ contains the corresponding MITRE ATT\&CK tactics and techniques \cite{mitre-attack}. Tokenizing each label as $\mathbf{y}^i = (y_1^i, \hdots, y^i_l)$, we fine-tune the model by sampling mini-batches and minimizing the standard autoregressive cross-entropy loss, where $\Phi_w$ denotes the LLM with tunable model weights $w\in \mathbb R^d$.
\begin{equation}
\label{eq:loss}
L(w)=-\frac{1}{B}\sum_{i=1}^B\sum_{k=1}^{l_i} \log \Phi_{w}(y_k^i|\mathbf{x}^i, y_{1:k-1}^{i}).
\end{equation}

\noindent\textit{\textbf{Belief generation fine-tuning.}} Next, we apply the same fine-tuning method to adapt a version of the LLM for belief estimation. Specifically, given incident observations and previous actions, the model is trained to estimate the belief $b_t$ and an evidence summary $m_t$, which contains the observations that support $b_t$. The training data consists of instruction-answer pairs where the input describes the incident context and the output provides the corresponding belief $b_t$ and evidence summary $m_t$. This fine-tuning enables the agent to compress incident context (e.g., logs and alerts) into a compact belief representation $(b_t, m_t)$.

\noindent\textit{\textbf{Response action fine-tuning.}} Lastly, we apply the same fine-tuning method to adapt a version of the LLM to generate response actions conditioned on the current belief state. In this case, each training example provides the incident description, the current belief $(b_t, m_t)$, and the previous response action $a_{t-1}$ as input. The model is then trained to predict the next local action $a^{\ell}_t$, which can be verified in the digital twin, as detailed below.

\subsection{Digital Twin for Simulation and Emulation}
\begin{figure}
\begin{minipage}{0.4\linewidth}
\includegraphics[width=1\linewidth]{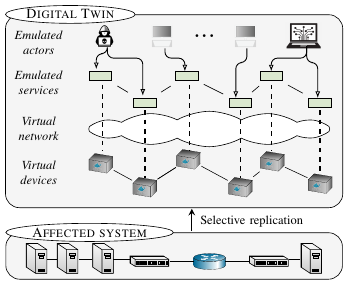}    
\end{minipage}
\begin{minipage}{0.6\linewidth}
\includegraphics[width=1\linewidth]{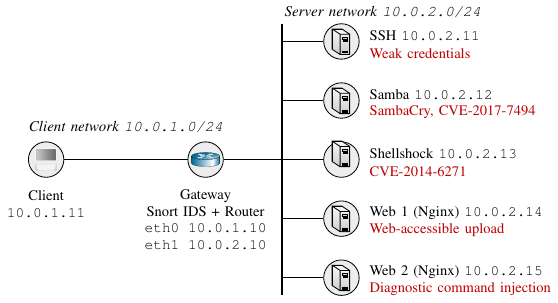} 
\end{minipage}
\caption{(Left) The digital twin is a virtual replica of the affected system, which offers a safe environment for investigating the incident and testing response actions. (Right) The network digital twin configuration adopted in our testbed.}
\label{fig:dt}
\vspace{-1em}
\end{figure}

The digital twin is an isolated execution environment that replicates the relevant hosts, services, and configurations of the system affected by the incident, as shown in Fig.~\ref{fig:dt}. It provides a safe environment to investigate the incident and evaluate response actions. For example, if the incident occurs in a cloud environment, then the digital twin can be created by taking snapshots of system components and deploying them in a private cloud. Without such a digital twin, investigation and testing must be performed directly on the affected system, which increases risks and operational costs.

In our method, the digital twin supports two complementary modes of execution: simulation and emulation. In simulation mode, it uses a network attack graph to evaluate attack progression and compare recovery-priority orders efficiently. This mode is used for tactical planning, where the goal is to decide which component to recover next. In emulation mode, the digital twin executes candidate recovery actions in a replica of the affected environment. This mode is used for operational planning, where the goal is to test whether generated response actions are effective before applying them to the operational system.

\noindent\textit{\textbf{Identification of an attack graph for simulation.}}
Before initiating tactical response planning, we use the digital twin and the previously fine-tuned LLM to identify an \textit{attack graph} for simulation. To this end, we prompt the LLM with the incident description $\mathbf{I}$ and use it to generate an initial assessment of the likely attack tactics and techniques. We refer to this estimate as an \textit{attack conjecture}. To improve the robustness of this conjecture, we run $M=10$ independent generations and retain only tactics and techniques with empirical frequency exceeding $0.5$. We denote the resulting conjecture by $\hat{\theta} \gets \Phi_w(\cdot \mid \mathbf{I})$.

We then use the digital twin to extract dependencies among system components, thereby defining a dependency graph $\mathcal{G}$. For each edge $i\rightarrow j\in \mathcal{G}$, we repeatedly test the attack techniques in $\hat{\theta}$ that could enable this attack step. Subsequently, we use the outcomes of these tests to estimate the probability of the attack step $i\rightarrow j$, and denote the resulting attack graph by $\mathcal{G}(\hat{\theta})$.

\subsection{Tactical and Operational Scale Planning}
\noindent\textit{\textbf{Tactical Planning.}} The attack graph $\mathcal{G}(\hat{\theta})$ induces a transition model $\hat{P}^g(g_{t+1} \mid g_t, a^g_t)$ of the global state, which we use for tactical scale planning to decide which system component should be recovered next. This tactical planning involves comparing candidate recovery orders through lookahead simulations based on the attack graph and then selecting the order that leads to the minimal cost, which we compute as 
$\hat{c}^g(g_t, a^g_t) \approx \lambda \left(\sum_{j\neq k}r_j(g_t)a^g_t(j)\right)\tau_{\text{avg}},$
where $k$ is the component placed first in the recovery order and $\tau_\text{avg}$ is the average local execution time of previously attended nodes. We use the average time since the exact recovery time is not known before selecting a component.

To evaluate candidate recovery orders, we first use the fine-tuned LLM to generate a belief state $b_t$. We then use this belief state to sample $M_g$ possible global states as $\{\hat{g}^i_t\}_{i \in [M_g]} \sim b^g_t$ and simulate $H_g$ lookahead steps in the attack graph. We then estimate the cost of the resulting rollout by averaging the sampled states as $J(\hat{a}^g) = \frac{1}{M_g}\sum_{i \in [M_g]}\sum_{k=0}^{H_g}\hat{c}^g(\hat{g}^i_{t+k}, \hat{a}^g)$, where $\hat{g}^i_{t+k}\sim\hat P^g(\cdot\mid \hat g^i_{t+k-1}, \hat a^g)$.
Next, we select the recovery order with the lowest estimated cost:
$
a^g_t = \argmin_{\hat{a} \in \mathcal{A}_t^g}J(\hat{a}).
$
Since enumerating all $N!$ recovery orders is computationally infeasible for large systems, we restrict this minimization to candidate orders derived by permuting the previous order $a^g_{t-1}$.

\noindent\textit{\textbf{Operational Planning.}}
Given the belief and system components selected by tactical planning, operational planning decides how to recover those components, i.e., it determines the next local recovery action $a^{\ell}_t$. This planning process involves three main steps. First, we use the fine-tuned LLM to generate a set of candidate response actions
$\mathcal{A}^{\ell}_t = \{\hat{a}^1_t, \hat{a}^2_t, \hdots, \hat{a}^{N_\ell}_t\}$.
Second, we evaluate the candidate actions using LLM-generated recovery rollouts. In particular, for each candidate action $\hat{a}^i_t$, we use the fine-tuned LLM to generate $M_{\ell} > 0$ recovery trajectories. Each trajectory starts with an action $\hat{a}^i_t$ and continues until the target component reaches the recovered state $\ell_T = (1,1,1,1,1,1)$. These rollouts are then used to estimate the remaining recovery cost as
\begin{align*}
Q(b_t, \hat{a}^i_t) = \frac{1}{M_{\ell}}\sum_{j \in [M_{\ell}]}\sum_{\hat{a} \in q^{i,j}}c(\hat{a}), c(\hat{a})=
\begin{dcases}
\operatorname{DT-Emul}(\hat a), & \text{if verified by DT},\\
\infty,  & \text{otherwise},
\end{dcases} 
\end{align*}
where $q^{i,j}$ is the $j$th rollout trajectory starting from action $\hat{a}^i_t$, and the time cost of each action is estimated by execution time in the digital twin, denoted by $\operatorname{DT-Emul}(\hat{a})$. If the action is not executable, the cost is set to $\infty$. 

The third step selects the action with the lowest estimated cost. 
We then execute the selected action in the digital twin. As a result of this execution, the digital twin produces a new observation $o_{t+1}$ (e.g., the outcome of the action), which we feed to the fine-tuned LLM to generate a new belief $(b_{t+1}, m_{t+1})$. We then repeat the same planning procedure from the new belief. This process of planning a local recovery action, executing it, and updating the belief continues until the component is recovered

\section{Experiment}
\label{sec:experiment}
In this section, we present an experimental evaluation of our method for agentic response planning. We start by assessing the effectiveness of offline fine-tuning. We then compare the recovery-action execution time and recovery rate of our method with frontier LLMs and two state-of-the-art baselines: \textsc{ircopilot} \cite{guo25ircopilot} and \textsc{llm-ir} \cite{kim26ndss}.

\subsection{Experiment Setup}\label{sec:experiment_setup}

We instantiate the LLM $\Phi_w$ [cf.~\eqref{eq:loss}] with \path{DeepSeek-R1-Distill-Qwen-14B} and fine-tune its weights using LoRA. The base-model parameters remain frozen during fine-tuning, while the trainable parameters are restricted to the LoRA adapter weights. The hyperparameters used for fine-tuning are available in Appendix~\ref{app:setup}. The prompt templates and related artifacts are available online\footnote{GitHub repository:\url{https://github.com/TaoLi-NYU/Agentic-Incident-Response-ESORICS26}}.

\noindent\textit{\textbf{Fine-tuning datasets.}}
We use the training dataset from \cite{kim26ndss} to fine-tune the LLM. This dataset is divided into three separate fine-tuning datasets: $\mathcal{D}=\mathcal{D}_{\mathrm{incident}}\cup\mathcal{D}_{\mathrm{state}}\cup\mathcal{D}_{\mathrm{action}}$ to fine-tune three versions of the LLM.

First, we use $\mathcal{D}_{\mathrm{incident}}$ for incident-assessment fine-tuning, where each training example consists of a system description and security logs. The model is fine-tuned to decide whether the evidence in the logs indicates an incident, summarize the incident, identify involved entities, and assign MITRE ATT\&CK tactics. 

Second, we use $\mathcal{D}_{\mathrm{state}}$ for belief-generation fine-tuning, where each training example includes the system description, logs, incident summary, local state, and previously executed response actions. The model is trained to predict the next global and local recovery state, along with a summary of evidence. The belief is obtained by the empirical distribution of repeatedly generated state predictions. 

Third, we use $\mathcal{D}_{\mathrm{action}}$, for action-generation fine-tuning, where each training example instructs the model to generate the next response action based on the local state and previous actions. 

\noindent\textit{\textbf{Digital twin.}}
We use a dockerized digital twin that emulates a small segmented enterprise network with two subnets: a client network \path{10.0.1.0/24} and a server network \path{10.0.2.0/24}; see Fig.~\ref{fig:dt}. The two subnets are connected by a gateway container, which acts as both the network router and the IDS monitoring point. The gateway has IP address \path{10.0.1.10} on the client network and \path{10.0.2.10} on the server network. It runs Snort to collect alerts and uses \path{iptables} to support containment and recovery actions. The client container has IP address \path{10.0.1.11} and serves as the attack platform in the experiments. It includes common network and exploitation tools, including \path{nmap}, \path{hydra}, \path{curl}, \path{smbclient}, and \path{sshpass}. The roles of the components are summarized in Table~\ref{tab:dt-testbed} in Appendix~\ref{app:setup}.

\noindent\textit{\textbf{Attack scenarios.}} We consider three different attack scenarios to evaluate our method. These scenarios, which we refer to as \textbf{Weak-Credential-3}, \textbf{Shellshock-4}, and \textbf{Command-Injection-5}, are designed to exhibit different combinations of attack patterns (e.g., command injection) that compromise different nodes in the digital twin. We defer the detailed setup to Appendix~\ref{app:setup}.

\subsection{LLM Generation Evaluation}
In this section, we present an evaluation of the generation capabilities of the fine-tuned LLMs. For each fine-tuned LLM, we evaluate the LLM on a set of testing incident examples reported in the literature, listed in \Cref{tab:dataset}.
\begin{table}[!ht]
    \centering
    \resizebox{\linewidth}{!}{
    \begin{tabular}{llll}
    \toprule
    Dataset & Systems & Attacks & logs\\
    \midrule
    CTU-Malware-2014 \cite{GARCIA2014100}    & \textsc{windows xp sp2} & Malwares and ransomwares & \textsc{snort} alerts\\
    CIC-IDS-2017 \cite{ghorbani18dataset}    & \textsc{windows}, \textsc{linux} & Denial-of-Service, web attacks, \textsc{sql} injections, and etc.&\textsc{snort} alerts\\
    AIT-IDS-V2-2022 \cite{Wurzenberger24} & \textsc{windows}, \textsc{linux} & Multi-stage attacks from reconnaissance to escalation& \textsc{wazuh} alerts\\
    CSLE-IDS-2024 \cite{kim24dataset} & \textsc{linux} & Software exploits, e.g., \textsc{cve}-2015-1427&\textsc{snort} alerts \\
    \bottomrule
    \end{tabular}
    }
    \caption{The four evaluation datasets used to evaluate the fine-tuned LLMs.}
    \label{tab:dataset}
    \vspace{-2em}
\end{table}

\noindent\textit{\textbf{Incident assessment evaluation.}} We begin by evaluating the fine-tuned LLM's ability to identify MITRE ATT\&CK tactics from incident descriptions and security logs. For each ground-truth tactic in the test data, we define precision as the fraction of predicted instances that correctly include the tactic among all predictions, and recall as the fraction of testing instances that contain the tactic and are correctly predicted. \Cref{tab:per-tactic} summarizes the results. We observe that the prediction accuracy is positively correlated with the occurrences of each tactic in the training dataset. Moreover, we find that the model achieves high precision and recall ($\geq 0.85$) for the most common tactics (e.g., reconnaissance), but lower accuracy for the less common tactics (e.g., discovery).
 \begin{table}
\begin{minipage}[t]{0.4\linewidth}
\vspace{0pt}
    \centering
    \resizebox{\linewidth}{!}{
    \begin{tabular}{lccc}
    \toprule
    Tactic & Precision & Recall & F1 \\
    \midrule
    Reconnaissance & 1.000 & 0.857 & 0.923 \\
    Initial Access & 0.839 & 0.975 & 0.902 \\
    Execution & 0.938 & 0.974 & 0.955 \\
    Persistence & 0.500 & 0.667 & 0.571 \\
    Privilege Escalation & 0.378 & 0.583 & 0.459 \\
    Defense Evasion & 0.417 & 0.714 & 0.526 \\
    Credential Access & 0.733 & 0.647 & 0.688 \\
    Discovery & 0.333 & 0.765 & 0.464 \\
    Lateral Movement & 0.527 & 0.879 & 0.659 \\
    Collection & 0.571 & 0.800 & 0.667 \\
    Command and Control & 0.690 & 0.925 & 0.790 \\
    Exfiltration & 0.833 & 0.893 & 0.862 \\
    Impact & 0.818 & 0.818 & 0.818 \\
    \bottomrule
    \end{tabular}}
    \caption{Evaluation of the LLM for incident assessment.}
    \label{tab:per-tactic}
\end{minipage}
\begin{minipage}[t]{0.58\linewidth}
\vspace{0pt}
\resizebox{\linewidth}{!}{
\begin{tabular}{lcccccccc}
      \toprule
          & caa-F1 & csa-F1 & $\ell^c$ & $\ell^a$ & $\ell^p$ & $\ell^e$ & $\ell^h$ & $\ell^r$ \\
      \midrule
      F1  & 0.9902 & 0.9822 & 0.9975 & 0.9964 & 0.9970 & 0.9952 & 0.9541 & 0.9533 \\
      \bottomrule
      \end{tabular}}
      \caption{Evaluation of the LLM for belief generation; reproduced from Tab. 2 \cite{yiran26e2e-ir}.}
      \label{tab:belief-f1}

    \vspace{2em}
    \resizebox{\linewidth}{!}{
    \begin{tabular}{lccc}
    \toprule
    Scenario & Ground truth  & Predicted belief  & BCE \\
    \midrule
    Weak-Credential-3 & $(0,0,0,1,1)$ & $(0.04,0.06,0.08,0.58,0.58)$ & 0.2551 \\
    Shellshock-4 & $(0,0,0,0,1)$ & $(0.04,0.08,0.04,0.10,0.92)$ & 0.0708 \\
    Command-Injection-5 & $(0,0,0,0,0)$ & $(0.10,0.12,0.10,0.26,0.10)$ & $0.1490$ \\
    \bottomrule
    \end{tabular}}
    \caption{Evaluation of the LLM for belief generation.}
    \label{tab:global-state-bce}
\end{minipage}
\vspace{-2em}
\end{table} 

\noindent\textit{\textbf{Belief generation evaluation.}} Since each local recovery state is a six-dim Boolean vector (see \Cref{sec:pre}), predicting the local state is a multi-label binary classification problem. For this reason, we evaluate the LLM's predictions using the F1 score. In addition to the F1 score, we also measure two averaged F1 scores: (\textit{i}) {class-agnostic average F1} (caa-F1), which aggregates True Positive (TP)/ False Positive (FP)/ False Negative (FN) across all entries before computing F1; and (\textit{ii}) {class-specific average F1} (csa-F1), which computes F1 per Boolean entry and then averages across entries. The F1 scores are summarized in \Cref{tab:belief-f1}. We find that our lightweight LLM can accurately predict the recovery state. 

In addition to the evaluation in terms of the F1 score of the state prediction, we consider the LLM-generated global state beliefs specific to the three attack scenarios in the digital twin. Specifically, \Cref{tab:global-state-bce} lists the binary cross-entropy between the generated belief and the true state. We observe in the table that the generated beliefs put most probability mass on the true state.

\noindent\textit{\textbf{Response action generation evaluation.}}
We then evaluate the response actions generated by the fine-tuned LLM for the five most frequent MITRE ATT\&CK tactics in the testing dataset. For each tactic, we randomly select 30 incident examples and prompt the model to generate actions, whose effectiveness is judged by \textsc{codex}, a \textsc{gpt}-powered coding agent with access to the ground-truth actions. A generated action is said to \textit{pass} the test if it advances the recovery progress to the same extent as the ground-truth action. \Cref{tab:action-recovery-effect} summarizes the pass rate across the 150 testing examples, indicating that our model can generate effective actions for different tactics.
\begin{table}[h]
\vspace{-1em}
    \centering
    \resizebox{\linewidth}{!}{
    \begin{tabular}
    {p{0.75\textwidth}lll}
    \toprule
    MITRE ATT\&CK tactics & \# & Pass & Fail \\
    \midrule
    Initial Access, Execution, Collection, Exfiltration & 30 & 24 (80.0\%) & 6 (20.0\%) \\
    Initial Access, Execution, Command and Control, Exfiltration & 30 & 23 (76.7\%) & 7 (23.3\%) \\
    Initial Access, Execution, Credential Access, Exfiltration  & 30 & 28 (93.3\%) & 2 (6.7\%) \\
    Impact & 30 & 25 (83.3\%) & 5 (16.7\%) \\
    Initial Access, Execution, Command and Control & 30 & 28 (93.3\%) & 2 (6.7\%) \\
    \midrule
    \textbf{Overall} & \textbf{150} & \textbf{128 (85.3\%)} &
    \textbf{22 (14.7\%)} \\
    \bottomrule
    \end{tabular}
    }
    \caption{Evaluation of the LLM for action generation.}
    \label{tab:action-recovery-effect}
    \vspace{-4em}
\end{table}

\subsection{Recovery-Action Execution-Time Evaluation}
In this section, we evaluate the execution performance of our agentic method using response action execution time (s) and recovery rate (\%). Execution time excludes planning, LLM inference, and DT-verification.  We consider two evaluation metrics: the recovery time (s) and the recovery rate (\%). The former measures the average time to recover the system from an incident using the planned sequence, and the latter measures the percentage of fully recovered incidents among those evaluated. We highlight that the recovery time excludes the planning time, LLM inference time, and DT verification time. All evaluations are performed using the digital twin shown in Fig.~\ref{fig:dt} and the incidents (attack scenarios) described in Appendix~\ref{app:setup}.

\noindent\textit{\textbf{Baselines.}} We use six baselines. The first four baselines are the current frontier LLMs, namely: \textsc{gpt-5.5} \cite{gpt5-5}, \textsc{gemini-3.1-pro} \cite{gemini3-1}, \textsc{deepseek-v4-pro} \cite{ds-v4}, and \textsc{claude-opus-4.8} (high effort) \cite{opus4-8}. Compared with these LLMs, our method is significantly more lightweight and fine-tuned towards the incident response use cases. In addition to the frontier LLMs, we compare the performance of our method with that of \textsc{ircopilot} \cite{guo25ircopilot} and \textsc{llm-ir} \cite{kim26ndss}, both of which are agentic response methods proposed in the literature. Compared with our agent, \textsc{ircopilot} uses our fine-tuned model only for operational planning, and \textsc{llm-ir} uses LLM-based rollout at the operational scale without digital twin verification. The two baselines constitute ablations of the usage of rollout and digital twin.

\noindent\textit{\textbf{Evaluation scenarios.}} We consider two evaluation scenarios. In the first scenario, we use the tactical-scale plan produced by our method to evaluate all baselines. Hence, this evaluation scenario compares the operational-scale planning capabilities of the different methods. In the second scenario, we measure the benefit of the tactical planning component of our method by prompting each baseline method to perform tactical-scale planning as well. 

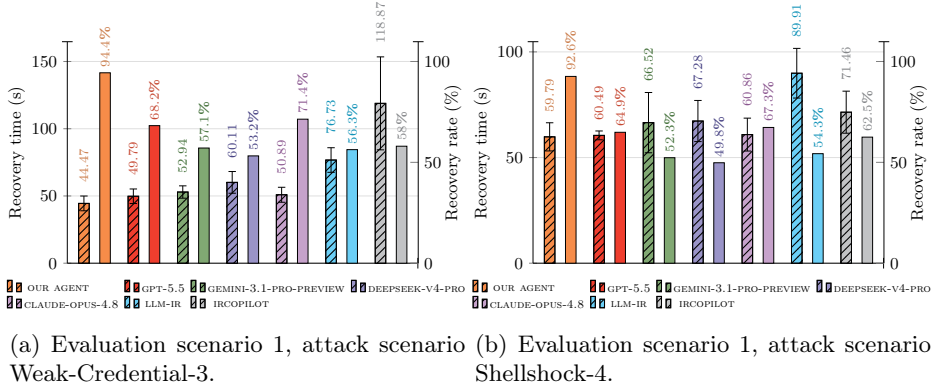
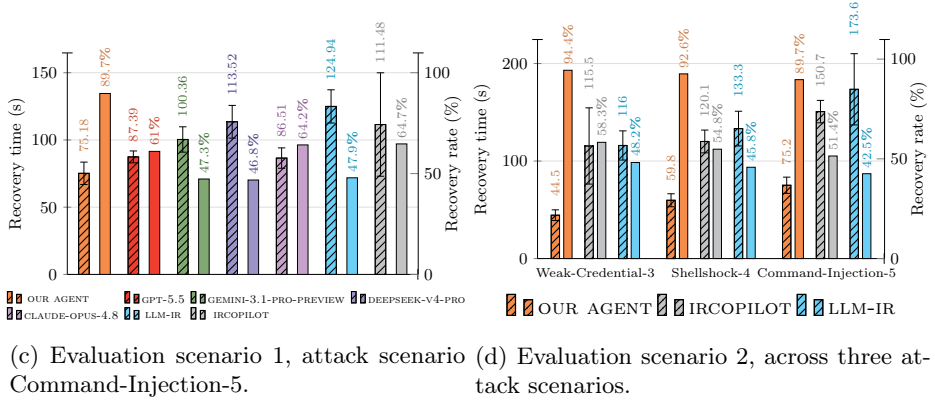
\begin{figure}[!ht]
\vspace{-1em}
\centering
\begin{subfigure}{0.495\linewidth}
\centering
\makebox[\linewidth][c]{%
  \resizebox{1.03\linewidth}{!}{\begin{tikzpicture}
\begin{axis}[
name=leftaxis,
scale only axis,
ybar,
ymin=0,
ymax=165,
ylabel={Recovery time (s)},
ylabel style={yshift=-0.1cm},
axis x line=bottom,
axis y line=left,
axis line style={-|},
ymajorgrids,
grid style={gray!25},
width=1.10\linewidth,
height=4.2cm,
bar width=0.20cm,
xmin=-0.55,
xmax=6.55,
xtick=\empty,
clip=false,
nodes near coords,
every node near coord/.append style={
anchor=west,
rotate=90,
yshift=0pt,
xshift=0pt,
font=\scriptsize\bfseries
},
legend to name=attackthreelegend,
legend style={
nodes={scale=0.85, transform shape},
at={(0.5,-0.10)},
anchor=north,
legend columns=4,
draw=none,
/tikz/every even column/.append style={column sep=0.08cm}
},
]

\addplot+[
draw=black,
fill=Orange!90,
postaction={pattern=north east lines},
bar shift=-0.20cm,
every node near coord/.append style={
xshift=8pt,
text=Orange!80!black
},
error bars/.cd,
error bar style={draw=black},
error mark options={rotate=90,draw=black,fill=black},
y dir=both,
y explicit,
] coordinates {
(0,44.473) +- (0,5.493)
};

\addplot+[
draw=black,
fill=Red!90,
postaction={pattern=north east lines},
bar shift=-0.20cm,
every node near coord/.append style={
xshift=8pt,
text=Red!80!black
},
error bars/.cd,
error bar style={draw=black},
error mark options={rotate=90,draw=black,fill=black},
y dir=both,
y explicit,
] coordinates {
(1,49.785) +- (0,5.414)
};

\addplot+[
draw=black,
fill=OliveGreen!60,
postaction={pattern=north east lines},
bar shift=-0.20cm,
every node near coord/.append style={
xshift=7pt,
text=OliveGreen!80!black
},
error bars/.cd,
error bar style={draw=black},
error mark options={rotate=90,draw=black,fill=black},
y dir=both,
y explicit,
] coordinates {
(2,52.937) +- (0,4.589)
};

\addplot+[
draw=black,
fill=Blue!40,
postaction={pattern=north east lines},
bar shift=-0.20cm,
every node near coord/.append style={
xshift=10pt,
text=Blue!75!black
},
error bars/.cd,
error bar style={draw=black},
error mark options={rotate=90,draw=black,fill=black},
y dir=both,
y explicit,
] coordinates {
(3,60.109) +- (0,8.076)
};

\addplot+[
draw=black,
fill=Purple!45,
postaction={pattern=north east lines},
bar shift=-0.20cm,
every node near coord/.append style={
xshift=8pt,
text=Purple!80!black
},
error bars/.cd,
error bar style={draw=black},
error mark options={rotate=90,draw=black,fill=black},
y dir=both,
y explicit,
] coordinates {
(4,50.889) +- (0,5.572)
};

\addplot+[
draw=black,
fill=Cyan!50,
postaction={pattern=north east lines},
bar shift=-0.20cm,
every node near coord/.append style={xshift=11pt,text=Cyan!70!black},
error bars/.cd,
error bar style={draw=black},
error mark options={rotate=90,draw=black,fill=black},
y dir=both,
y explicit,
] coordinates {
(5,76.726) +- (0,9.148)
};

\addplot+[
draw=black,
fill=Gray!55,
postaction={pattern=north east lines},
bar shift=-0.20cm,
every node near coord/.append style={text=Gray!85!black, xshift=29pt},
error bars/.cd,
error bar style={draw=black},
error mark options={rotate=90,draw=black,fill=black},
y dir=both,
y explicit,
] coordinates {
(6,118.871) +- (0,34.569)
};

\legend{
\hspace{-0.7cm}\textsc{our agent},
\textsc{gpt-5.5},
\textsc{gemini-3.1-pro-preview},
\textsc{deepseek-v4-pro},
\textsc{claude-opus-4.8},
\textsc{llm-ir},
\hspace{-1.9cm}\textsc{ircopilot}
}

\end{axis}

\begin{axis}[
at={(leftaxis.south west)},
anchor=south west,
scale only axis,
ybar,
ymin=0,
ymax=110,
ylabel={Recovery rate (\%)},
ylabel style={
at={(axis description cs:1.14,0.5)},
anchor=south
},
axis x line=none,
axis y line*=right,
axis line style={-|},
width=1.10\linewidth,
height=4.2cm,
bar width=0.20cm,
xmin=-0.55,
xmax=6.55,
xtick=\empty,
clip=false,
point meta=y,
nodes near coords={
\pgfmathprintnumber[fixed,precision=1]{\pgfplotspointmeta}\%
},
every node near coord/.append style={
anchor=west,
rotate=90,
yshift=0pt,
xshift=0pt,
font=\scriptsize\bfseries
},
]

\addplot+[
draw=black,
fill=Orange!90,
bar shift=0.20cm,
every node near coord/.append style={text=Orange!80!black}
] coordinates {(0,94.4)};

\addplot+[
draw=black,
fill=Red!90,
bar shift=0.20cm,
every node near coord/.append style={text=Red!80!black}
] coordinates {(1,68.2)};

\addplot+[
draw=black,
fill=OliveGreen!60,
bar shift=0.20cm,
every node near coord/.append style={text=OliveGreen!80!black}
] coordinates {(2,57.1)};

\addplot+[
draw=black,
fill=Blue!40,
bar shift=0.20cm,
every node near coord/.append style={text=Blue!75!black}
] coordinates {(3,53.2)};

\addplot+[
draw=black,
fill=Purple!45,
bar shift=0.20cm,
every node near coord/.append style={text=Purple!80!black}
] coordinates {(4,71.4)};

\addplot+[
draw=black,
fill=Cyan!50,
bar shift=0.20cm,
every node near coord/.append style={text=Cyan!70!black}
] coordinates {(5,56.3)};

\addplot+[
draw=black,
fill=Gray!55,
bar shift=0.20cm,
every node near coord/.append style={text=Gray!85!black}
] coordinates {(6,58.0)};

\end{axis}
\end{tikzpicture}}%
}
\par
\makebox[\linewidth][c]{%
  \resizebox{1.03\linewidth}{!}{\pgfplotslegendfromname{attackthreelegend}}%
}
\caption{Evaluation scenario 1, attack scenario Weak-Credential-3. }
\label{fig:recovery-time-3}    
\end{subfigure}\hfill%
\begin{subfigure}{0.495\linewidth}
\centering
\makebox[\linewidth][c]{%
  \resizebox{1.03\linewidth}{!}{\begin{tikzpicture}
\begin{axis}[
name=leftaxis,
scale only axis,
ybar,
ymin=0,
ymax=105,
ylabel={Recovery time (s)},
ylabel style={yshift=-0.1cm},
axis x line=bottom,
axis y line=left,
axis line style={-|},
ymajorgrids,
grid style={gray!25},
width=1.10\linewidth,
height=4.2cm,
bar width=0.20cm,
xmin=-0.55,
xmax=6.55,
xtick=\empty,
clip=false,
nodes near coords,
every node near coord/.append style={
anchor=west,
rotate=90,
yshift=0pt,
xshift=0pt,
font=\scriptsize\bfseries
},
legend to name=attackfourlegend,
legend style={
nodes={scale=0.85, transform shape},
at={(0.5,-0.10)},
anchor=north,
legend columns=4,
draw=none,
/tikz/every even column/.append style={column sep=0.08cm}
},
]

\addplot+[
draw=black,
fill=Orange!90,
postaction={pattern=north east lines},
bar shift=-0.20cm,
every node near coord/.append style={xshift=10pt,text=Orange!80!black},
error bars/.cd,
error bar style={draw=black},
error mark options={rotate=90,draw=black,fill=black},
y dir=both,
y explicit,
] coordinates {
(0,59.791) +- (0,6.668)
};

\addplot+[
draw=black,
fill=Red!90,
postaction={pattern=north east lines},
bar shift=-0.20cm,
every node near coord/.append style={xshift=5pt,text=Red!80!black},
error bars/.cd,
error bar style={draw=black},
error mark options={rotate=90,draw=black,fill=black},
y dir=both,
y explicit,
] coordinates {
(1,60.487) +- (0,2.091)
};

\addplot+[
draw=black,
fill=OliveGreen!60,
postaction={pattern=north east lines},
bar shift=-0.20cm,
every node near coord/.append style={xshift=19pt,text=OliveGreen!80!black},
error bars/.cd,
error bar style={draw=black},
error mark options={rotate=90,draw=black,fill=black},
y dir=both,
y explicit,
] coordinates {
(2,66.519) +- (0,14.258)
};

\addplot+[
draw=black,
fill=Blue!40,
postaction={pattern=north east lines},
bar shift=-0.20cm,
every node near coord/.append style={xshift=14pt,text=Blue!75!black},
error bars/.cd,
error bar style={draw=black},
error mark options={rotate=90,draw=black,fill=black},
y dir=both,
y explicit,
] coordinates {
(3,67.280) +- (0,9.735)
};

\addplot+[
draw=black,
fill=Purple!45,
postaction={pattern=north east lines},
bar shift=-0.20cm,
every node near coord/.append style={xshift=12pt,text=Purple!80!black},
error bars/.cd,
error bar style={draw=black},
error mark options={rotate=90,draw=black,fill=black},
y dir=both,
y explicit,
] coordinates {
(4,60.863) +- (0,7.784)
};

\addplot+[
draw=black,
fill=Cyan!50,
postaction={pattern=north east lines},
bar shift=-0.20cm,
every node near coord/.append style={xshift=16pt,text=Cyan!70!black},
error bars/.cd,
error bar style={draw=black},
error mark options={rotate=90,draw=black,fill=black},
y dir=both,
y explicit,
] coordinates {
(5,89.907) +- (0,11.750)
};

\addplot+[
draw=black,
fill=Gray!55,
postaction={pattern=north east lines},
bar shift=-0.20cm,
every node near coord/.append style={xshift=14pt,text=Gray!85!black},
error bars/.cd,
error bar style={draw=black},
error mark options={rotate=90,draw=black,fill=black},
y dir=both,
y explicit,
] coordinates {
(6,71.463) +- (0,9.951)
};

\legend{
\hspace{-0.7cm}\textsc{our agent},
\textsc{gpt-5.5},
\textsc{gemini-3.1-pro-preview},
\textsc{deepseek-v4-pro},
\textsc{claude-opus-4.8},
\textsc{llm-ir},
\hspace{-1.9cm}\textsc{ircopilot}
}

\end{axis}

\begin{axis}[
at={(leftaxis.south west)},
anchor=south west,
scale only axis,
ybar,
ymin=0,
ymax=110,
ylabel={Recovery rate (\%)},
ylabel style={
at={(axis description cs:1.14,0.5)},
anchor=south
},
axis x line=none,
axis y line*=right,
axis line style={-|},
width=1.10\linewidth,
height=4.2cm,
bar width=0.20cm,
xmin=-0.55,
xmax=6.55,
xtick=\empty,
clip=false,
point meta=y,
nodes near coords={
\pgfmathprintnumber[fixed,precision=1]{\pgfplotspointmeta}\%
},
every node near coord/.append style={
anchor=west,
rotate=90,
yshift=0pt,
xshift=0pt,
font=\scriptsize\bfseries
},
]

\addplot+[
draw=black,
fill=Orange!90,
bar shift=0.20cm,
every node near coord/.append style={text=Orange!80!black}
]
coordinates {(0,92.6)};

\addplot+[
draw=black,
fill=Red!90,
bar shift=0.20cm,
every node near coord/.append style={text=Red!80!black}
]
coordinates {(1,64.9)};

\addplot+[
draw=black,
fill=OliveGreen!60,
bar shift=0.20cm,
every node near coord/.append style={text=OliveGreen!80!black}
]
coordinates {(2,52.3)};

\addplot+[
draw=black,
fill=Blue!40,
bar shift=0.20cm,
every node near coord/.append style={text=Blue!75!black}
]
coordinates {(3,49.8)};

\addplot+[
draw=black,
fill=Purple!45,
bar shift=0.20cm,
every node near coord/.append style={text=Purple!80!black}
]
coordinates {(4,67.3)};

\addplot+[
draw=black,
fill=Cyan!50,
bar shift=0.20cm,
every node near coord/.append style={text=Cyan!70!black}
]
coordinates {(5,54.3)};

\addplot+[
draw=black,
fill=Gray!55,
bar shift=0.20cm,
every node near coord/.append style={text=Gray!85!black}
]
coordinates {(6,62.5)};

\end{axis}
\end{tikzpicture}}%
}
\par
\makebox[\linewidth][c]{%
  \resizebox{1.03\linewidth}{!}{\pgfplotslegendfromname{attackfourlegend}}%
}
\caption{Evaluation scenario 1, attack scenario Shellshock-4.}
\label{fig:recovery-time-4}    
\end{subfigure}
\par\medskip

\begin{subfigure}[t]{0.495\linewidth}
\centering
\makebox[\linewidth][c]{%
  \resizebox{1.03\linewidth}{!}{\begin{tikzpicture}
\begin{axis}[
name=leftaxis,
scale only axis,
ybar,
ymin=0,
ymax=165,
ylabel={Recovery time (s)},
ylabel style={yshift=-0.1cm},
axis x line=bottom,
axis y line=left,
axis line style={-|},
ymajorgrids,
grid style={gray!25},
width=1.10\linewidth,
height=4.2cm,
bar width=0.20cm,
xmin=-0.55,
xmax=6.55,
xtick=\empty,
clip=false,
nodes near coords,
every node near coord/.append style={
anchor=west,
rotate=90,
yshift=0pt,
xshift=0pt,
font=\scriptsize\bfseries
},
legend to name=attackfivelegend,
legend style={
nodes={scale=0.85, transform shape},
at={(0.5,-0.10)},
anchor=north,
legend columns=4,
draw=none,
/tikz/every even column/.append style={column sep=0.08cm}
},
]

\addplot+[
draw=black,
fill=Orange!90,
postaction={pattern=north east lines},
bar shift=-0.20cm,
every node near coord/.append style={xshift=8pt,text=Orange!80!black},
error bars/.cd,
error bar style={draw=black},
error mark options={rotate=90,draw=black,fill=black},
y dir=both,
y explicit,
] coordinates {
(0,75.184) +- (0,8.33)
};

\addplot+[
draw=black,
fill=Red!90,
postaction={pattern=north east lines},
bar shift=-0.20cm,
every node near coord/.append style={xshift=5pt,text=Red!80!black},
error bars/.cd,
error bar style={draw=black},
error mark options={rotate=90,draw=black,fill=black},
y dir=both,
y explicit,
] coordinates {
(1,87.391) +- (0,4.425)
};

\addplot+[
draw=black,
fill=OliveGreen!60,
postaction={pattern=north east lines},
bar shift=-0.20cm,
every node near coord/.append style={xshift=9pt,text=OliveGreen!80!black},
error bars/.cd,
error bar style={draw=black},
error mark options={rotate=90,draw=black,fill=black},
y dir=both,
y explicit,
] coordinates {
(2,100.36) +- (0,9.4235)
};

\addplot+[
draw=black,
fill=Blue!40,
postaction={pattern=north east lines},
bar shift=-0.20cm,
every node near coord/.append style={xshift=11pt,text=Blue!75!black},
error bars/.cd,
error bar style={draw=black},
error mark options={rotate=90,draw=black,fill=black},
y dir=both,
y explicit,
] coordinates {
(3,113.516) +- (0,12.168)
};

\addplot+[
draw=black,
fill=Purple!45,
postaction={pattern=north east lines},
bar shift=-0.20cm,
every node near coord/.append style={xshift=8pt,text=Purple!80!black},
error bars/.cd,
error bar style={draw=black},
error mark options={rotate=90,draw=black,fill=black},
y dir=both,
y explicit,
] coordinates {
(4,86.513) +- (0,7.578)
};

\addplot+[
draw=black,
fill=Cyan!50,
postaction={pattern=north east lines},
bar shift=-0.20cm,
every node near coord/.append style={xshift=11pt, text=Cyan!70!black},
error bars/.cd,
error bar style={draw=black},
error mark options={rotate=90,draw=black,fill=black},
y dir=both,
y explicit,
] coordinates {
(5,124.940) +- (0,12.360)
};

\addplot+[
draw=black,
fill=Gray!55,
postaction={pattern=north east lines},
bar shift=-0.20cm,
every node near coord/.append style={xshift=30pt, text=Gray!85!black},
error bars/.cd,
error bar style={draw=black},
error mark options={rotate=90,draw=black,fill=black},
y dir=both,
y explicit,
] coordinates {
(6,111.478) +- (0,38.465)
};

\legend{
\hspace{-0.7cm}\textsc{our agent},
\textsc{gpt-5.5},
\textsc{gemini-3.1-pro-preview},
\textsc{deepseek-v4-pro},
\textsc{claude-opus-4.8},
\textsc{llm-ir},
\hspace{-1.9cm}\textsc{ircopilot}
}

\end{axis}

\begin{axis}[
at={(leftaxis.south west)},
anchor=south west,
scale only axis,
ybar,
ymin=0,
ymax=110,
ylabel={Recovery rate (\%)},
ylabel style={
at={(axis description cs:1.14,0.5)},
anchor=south
},
axis x line=none,
axis y line*=right,
axis line style={-|},
width=1.10\linewidth,
height=4.2cm,
bar width=0.20cm,
xmin=-0.55,
xmax=6.55,
xtick=\empty,
clip=false,
point meta=y,
nodes near coords={
\pgfmathprintnumber[fixed,precision=1]{\pgfplotspointmeta}\%
},
every node near coord/.append style={
anchor=west,
rotate=90,
yshift=0pt,
xshift=0pt,
font=\scriptsize\bfseries
},
]

\addplot+[
draw=black,
fill=Orange!90,
bar shift=0.20cm,
every node near coord/.append style={text=Orange!80!black}
]
coordinates {(0,89.7)};
\addplot+[
draw=black,
fill=Red!90,
bar shift=0.20cm,
every node near coord/.append style={text=Red!80!black}
]
coordinates {(1,61.0)};
\addplot+[
draw=black,
fill=OliveGreen!60,
bar shift=0.20cm,
every node near coord/.append style={text=OliveGreen!80!black}
]
coordinates {(2,47.3)};
\addplot+[
draw=black,
fill=Blue!40,
bar shift=0.20cm,
every node near coord/.append style={text=Blue!75!black}
]
coordinates {(3,46.8)};
\addplot+[
draw=black,
fill=Purple!45,
bar shift=0.20cm,
every node near coord/.append style={text=Purple!80!black}
]
coordinates {(4,64.2)};
\addplot+[
draw=black,
fill=Cyan!50,
bar shift=0.20cm,
every node near coord/.append style={text=Cyan!70!black}
]
coordinates {(5,47.9)};
\addplot+[
draw=black,
fill=Gray!55,
bar shift=0.20cm,
every node near coord/.append style={text=Gray!85!black}
]
coordinates {(6,64.7)};

\end{axis}
\end{tikzpicture}}%
}
\par
\makebox[\linewidth][c]{%
  \resizebox{1.03\linewidth}{!}{\pgfplotslegendfromname{attackfivelegend}}%
}
\caption{Evaluation scenario 1, attack scenario Command-Injection-5.}
\label{fig:recovery-time-5}  
\end{subfigure}\hfill%
\begin{subfigure}[t]{0.495\linewidth}
\centering
\makebox[\linewidth][c]{%
  \resizebox{1.03\linewidth}{!}{\begin{tikzpicture}
\begin{axis}[
name=systemleftaxis,
scale only axis,
ybar,
ymin=0,
ymax=225,
ylabel={Recovery time (s)},
axis x line=bottom,
axis y line=left,
axis line style={-|},
ymajorgrids,
grid style={gray!25},
width=1.10\linewidth,
height=4.2cm,
bar width=0.16cm,
xmin=-0.25,
xmax=3.35,
xtick={0.35,1.55,2.75},
xticklabels={Weak-Credential-3,Shellshock-4,Command-Injection-5},
xtick style={draw=none},
x tick label style={font=\scriptsize,align=center},
clip=false,
point meta=y,
nodes near coords={
  \pgfmathprintnumber[fixed,precision=1]{\pgfplotspointmeta}
},
every node near coord/.append style={
  anchor=west,
  rotate=90,
  yshift=0pt,
  xshift=0pt,
  font=\scriptsize\bfseries
},
legend to name=attacksystemlegend,
legend style={
  nodes={scale=0.85, transform shape},
  legend columns=3,
  draw=none
},
]

\addplot+[
draw=black,
fill=Orange!90,
postaction={pattern=north east lines},
bar shift=-0.12cm,
every node near coord/.append style={xshift=7pt,text=Orange!80!black},
error bars/.cd,
error bar style={draw=black},
error mark options={rotate=90,mark size=2pt,draw=black,fill=black},
y dir=both,
y explicit,
] coordinates {
(0.00,44.473) +- (0,5.493)
(1.20,59.791) +- (0,6.668)
};

\addplot+[
draw=black,
fill=Orange!90,
postaction={pattern=north east lines},
bar shift=-0.12cm,
forget plot,
every node near coord/.append style={
  text=Orange!80!black,
  xshift=8pt
},
error bars/.cd,
error bar style={draw=black},
error mark options={rotate=90,mark size=2pt,draw=black,fill=black},
y dir=both,
y explicit,
] coordinates {
(2.40,75.184) +- (0,8.330)
};

\addplot+[
draw=black,
fill=Gray!55,
postaction={pattern=north east lines},
bar shift=-0.12cm,
forget plot,
every node near coord/.append style={
  text=Gray!85!black,
  xshift=27pt
},
error bars/.cd,
error bar style={draw=black},
error mark options={rotate=90,mark size=2pt,draw=black,fill=black},
y dir=both,
y explicit,
] coordinates {
(0.35,115.470) +- (0,39.097)
};

\addplot+[
draw=black,
fill=Gray!55,
postaction={pattern=north east lines},
bar shift=-0.12cm,
forget plot,
every node near coord/.append style={
  text=Gray!85!black,
  xshift=10pt
},
error bars/.cd,
error bar style={draw=black},
error mark options={rotate=90,mark size=2pt,draw=black,fill=black},
y dir=both,
y explicit,
] coordinates {
(1.55,120.094) +- (0,11.671)
};

\addplot+[
draw=black,
fill=Gray!55,
postaction={pattern=north east lines},
bar shift=-0.12cm,
every node near coord/.append style={
  text=Gray!85!black,
  xshift=10pt
},
error bars/.cd,
error bar style={draw=black},
error mark options={rotate=90,mark size=2pt,draw=black,fill=black},
y dir=both,
y explicit,
] coordinates {
(2.75,150.691) +- (0,11.507)
};

\addplot+[
draw=black,
fill=Cyan!50,
postaction={pattern=north east lines},
bar shift=-0.12cm,
forget plot,
every node near coord/.append style={
  text=Cyan!70!black,
  xshift=12pt
},
error bars/.cd,
error bar style={draw=black},
error mark options={rotate=90,mark size=2pt,draw=black,fill=black},
y dir=both,
y explicit,
] coordinates {
(0.70,115.994) +- (0,14.970)
};

\addplot+[
draw=black,
fill=Cyan!50,
postaction={pattern=north east lines},
bar shift=-0.12cm,
forget plot,
every node near coord/.append style={
  text=Cyan!70!black,
  xshift=14pt
},
error bars/.cd,
error bar style={draw=black},
error mark options={rotate=90,mark size=2pt,draw=black,fill=black},
y dir=both,
y explicit,
] coordinates {
(1.90,133.276) +- (0,17.791)
};

\addplot+[
draw=black,
fill=Cyan!50,
postaction={pattern=north east lines},
bar shift=-0.12cm,
every node near coord/.append style={
  text=Cyan!70!black,
  xshift=26pt
},
error bars/.cd,
error bar style={draw=black},
error mark options={rotate=90,mark size=2pt,draw=black,fill=black},
y dir=both,
y explicit,
] coordinates {
(3.10,173.573) +- (0,36.542)
};

\legend{
\textsc{our agent},
\textsc{ircopilot},
\textsc{llm-ir}
}
\end{axis}

\begin{axis}[
at={(systemleftaxis.south west)},
anchor=south west,
scale only axis,
ybar,
ymin=0,
ymax=110,
ylabel={Recovery rate (\%)},
ylabel style={
  at={(axis description cs:1.14,0.5)},
  anchor=south
},
axis x line=none,
axis y line*=right,
axis line style={-|},
width=1.10\linewidth,
height=4.2cm,
bar width=0.16cm,
xmin=-0.25,
xmax=3.35,
xtick=\empty,
clip=false,
point meta=y,
nodes near coords={
  \pgfmathprintnumber[fixed,precision=1]{\pgfplotspointmeta}\%
},
every node near coord/.append style={
  anchor=west,
  rotate=90,
  yshift=0pt,
  xshift=0pt,
  font=\scriptsize\bfseries
},
]

\addplot+[
draw=black,
fill=Orange!90,
bar shift=0.12cm,
forget plot,
every node near coord/.append style={text=Orange!80!black}
] coordinates {
(0.00,94.4) (1.20,92.6) (2.40,89.7)
};

\addplot+[
draw=black,
fill=Gray!55,
bar shift=0.12cm,
forget plot,
every node near coord/.append style={text=Gray!85!black}
] coordinates {
(0.35,58.27) (1.55,54.83) (2.75,51.42)
};

\addplot+[
draw=black,
fill=Cyan!50,
bar shift=0.12cm,
forget plot,
every node near coord/.append style={text=Cyan!70!black}
] coordinates {
(0.70,48.16) (1.90,45.78) (3.10,42.54)
};

\end{axis}
\end{tikzpicture}}%
}
\par
\makebox[\linewidth][c]{%
  \resizebox{0.90\linewidth}{!}{\pgfplotslegendfromname{attacksystemlegend}}%
}
\caption{Evaluation scenario 2, across three attack scenarios.}
\label{fig:recovery-system-3-4-5}   
\end{subfigure}
\caption{Evaluation of recovery-action execution time and rate in the end-to-end incident response. The bars relate to the recovery time ($\downarrow$ better) and the recovery rate ($\uparrow$ better).}
\label{fig:time-rate}
\end{figure}

\noindent\textit{\textbf{Evaluation results.}} The results for evaluation scenario $1$ are summarized in Figs.~\ref{fig:recovery-time-3}, \ref{fig:recovery-time-4}, and \ref{fig:recovery-time-5}. Bar colors represent different methods; numbers and error bars indicate the mean and standard deviation from 50 independent runs using distinct random seeds. Notably, the recovery rates of our method, which are around 90\% in three attack scenarios, surpass all other baselines by a large margin. In contrast, the frontier models only achieve rates below 75\%, and the rates for two LLM-based approaches are between 50\% and 65\%. Moreover, our method achieves the shortest average recovery time, outperforming the frontier models by 15.1\%. We attribute this result to the fine-tuning.

The evaluation results related to scenario $2$ are summarized in Fig.~\ref{fig:recovery-system-3-4-5}. Similar to the results for scenario $1$, we see that our method maintains a recovery rate of around 90\% on average, which is above 30\% higher than the two baselines. The average recovery time of our method is 57.8\% and 53.9\% less than those of \textsc{llm-ir} and \textsc{ircopilot}, respectively. These results highlight the benefit of tactical planning.  

\subsection{Discussion}
The experimental results demonstrate that our method for agentic multiscale response planning consistently outperforms the frontier LLMs and prior works on LLM-based incident response across three attack scenarios. The comparisons with frontier LLMs suggest that a tailored, lightweight model powered by decision-theoretic planning provides a sufficient agentic approach for incident response. We attribute this result to the digital twin-based verification, which leads to higher recovery rates in our evaluation. Moreover, compared with prior LLM-based incident response approaches, the tactical-scale planning in our method improves performance by strategically selecting the recovery order.

\noindent\textit{\textbf{Limitations.}}
One limitation of our method is the misspecification of attack tactics during incident assessment. If the inferred attack conjecture deviates from the true tactics, the attack graph becomes inaccurate, and tactical-scale planning becomes less effective. While prior work has begun to address misspecification in security settings \cite{kim-tao25col,kim-tao25quantization}, this challenge remains unexplored when an LLM (i.e., a black-box generative model) serves as the inference model. Another limitation of our method is the cost of LLM-based rollout in operational planning. Failed verification adds extra cost due to regeneration and digital twin reinitialization. This can be partially mitigated through parallel LLM sessions and digital twin emulations.

\noindent\textit{\textbf{Connections to response playbooks.}} Our agentic response planning method serves a similar role to incident response playbooks \cite{applebaum}, but focuses on generating executable operational actions. Compared with conventional playbooks, it offers two advantages. First, it produces concrete and context-specific actions tailored to the system of interest. Second, it acts as a dynamic playbook by prioritizing compromised nodes through multiscale planning. 

\section{Conclusion}
We present an agentic approach that integrates decision-theoretic planning, large language models (LLMs), and digital twins for incident response planning. The proposed method uses a rollout planner based on digital twin simulation to compute a high-level response strategy that allocates security resources at the tactical scale. A lightweight LLM, fine-tuned over public incident datasets, translates the abstract strategies into executable commands verified by the digital twin emulation at the operational scale. We evaluate the agentic response planning approach on logs reported in the literature and three distinct attack scenarios on a testbed. Across three scenarios, our agentic approach maintains a recovery rate of around 90\%, reduces recovery-action execution time by 15.1\% on average, and increases the recovery rate by 33.6\% over the frontier LLMs.

A primary direction for future work is to address the attack tactics misspecification in the LLM-based incident assessment, which plays an instrumental role in tactical planning. The key is to leverage LLMs' in-context learning ability to calibrate their inference about potential tactics in the network by comparing observed alerts with predicted alerts under misspecified tactics. 
\bibliographystyle{splncs04}
\bibliography{ref}

\appendix
\section{Experiment Setup Details}
\label{app:setup}
Offline fine-tuning is performed on a Google Cloud virtual machine with 1 Nvidia A100 GPU, and the LoRA hyperparameters are presented in \Cref{tab:lora}. The components of our digital twin testbed are summarized in \Cref{tab:dt-testbed}.  
\begin{table}
\vspace{-1em}
\begin{minipage}[t]{0.37\linewidth}
\resizebox{\linewidth}{!}{
\begin{tabular}{ll}
      \toprule
      Parameter(s) & Value(s) \\
      \midrule
      LoRA rank, scaling, dropout & 64, 128, 0.05 \\
      Learning rate & $9.5\times 10^{-4}$ \\
      Per-device batch size & 1 \\
      Gradient accumulation steps & 32 \\
      Effective batch size & 32 \\
      Precision & bfloat16 \\
      \bottomrule
      \end{tabular}
      }
      \caption{A summary of LoRA fine-tuning hyperparameters.}
      \label{tab:lora}    
\end{minipage}
\begin{minipage}[t]{0.6\linewidth}
\resizebox{\linewidth}{!}{
 \begin{tabular}{lll}
      \toprule
      Host & IP address & Service role \\
      \midrule
      \path{server_ssh} & \path{10.0.2.11} & OpenSSH service \\
      \path{server_samba} & \path{10.0.2.12} & Samba file sharing \\
      \path{server_shellshock} & \path{10.0.2.13} & Apache CGI service \\
      \path{server_web1} & \path{10.0.2.14} & Nginx and upload service \\
      \path{server_web2} & \path{10.0.2.15} & Nginx and diagnostic service \\
      \bottomrule
      \end{tabular}
      }
      \caption{Components of the digital twin.}
      \label{tab:dt-testbed}    
\end{minipage}
\vspace{-2em}
\end{table}


\noindent\textit{\textbf{Attack scenarios.}} The first scenario, which we refer to as \textbf{Weak-Credential-3}, represents a multi-stage attack launched from the client host \path{10.0.1.11} that eventually takes down three nodes. The attack starts with network reconnaissance and TCP service scanning across the server network. It then compromises \path{server_ssh} at \path{10.0.2.11} through weak SSH credentials, accesses \path{server_samba} at \path{10.0.2.12} through an exposed anonymous SMB share, and exploits the Shellshock-vulnerable CGI service on \path{server_shellshock} at \path{10.0.2.13}. The attack also probes the two web servers at \path{10.0.2.14} and \path{10.0.2.15}. Finally, it uses the compromised SSH server as a pivot to reach other internal services. This scenario mainly compromises the first three servers in \Cref{tab:dt-testbed}.

The second scenario, referred to as \textbf{Shellshock-4}, targets four servers: \path{10.0.2.11}, \path{10.0.2.12}, \path{10.0.2.13}, and \path{10.0.2.14}. Unlike the first scenario, it avoids a broad ICMP sweep and instead performs targeted service discovery. It uses low-volume SSH password spraying against \path{server_ssh}, anonymous SMB share access against \path{server_samba}, Shellshock exploitation against \path{server_shellshock}, and unauthorized HTTP upload against \path{server_web1}. The scenario also includes a pivot from \path{server_shellshock}, where commands executed on \path{10.0.2.13} are used to probe reachable services on the other targeted servers.

The last scenario extends the attack coverage to all five servers in the server network, which we refer to as \textbf{Command-Injection-5}. This scenario increases attack techniques diversity by using a) service-specific TCP and HTTP fingerprinting; b) SSH credential reuse with file transfer; c) multi-file SMB staging and rename operations; d) Shellshock command execution; and e) command injection through the diagnostic service on \path{server_web2}. In contrast to the previous scenarios, the pivot host is \path{server_web2}. After command injection, the attack uses this host to probe reachable services on the other servers. This scenario evaluates the framework under a full five-server compromise with a different pivot point and a broader mixture of service-specific attack evidence.

\end{document}